\documentclass[%
 aps,
 amsmath,amssymb,
 reprint,%
]{revtex4-1}

\usepackage{graphicx}
\usepackage{dcolumn}
\usepackage{bm}

\usepackage[utf8]{inputenc}
\usepackage[T1]{fontenc}
\usepackage{mathptmx}
\usepackage{etoolbox}
\usepackage{hyperref}
\usepackage{float}
\usepackage[version=4]{mhchem}
\usepackage{multirow}
\usepackage{booktabs}
\usepackage[normalem]{ulem}
\usepackage{xcolor}
\usepackage{gensymb}

\newcommand{\newt}[1]{\textcolor{blue}{#1}}

\makeatletter
\def\@email#1#2{%
 \endgroup
 \patchcmd{\titleblock@produce}
  {\frontmatter@RRAPformat}
  {\frontmatter@RRAPformat{\produce@RRAP{*#1\href{mailto:#2}{#2}}}\frontmatter@RRAPformat}
  {}{}
}%
\makeatother
\begin{document}

\preprint{AIP/123-QED}

\title[Amorphous insulating MoS$_x$ films deposited by magnetron sputtering as a low-index optical coating]{Amorphous insulating MoS$_x$ films deposited by magnetron sputtering as a low-index optical coating}
\author{S.G. Martanov$^1$, E.V. Tarkaeva$^1$, A.V. Muratov$^1$, A.V. Lubenchenko$^2$, O.N. Pavlov$^2$, A.G. Vitukhnovsky$^3$, and A.Yu. Kuntsevich$^{1,*}$}
\affiliation{ 
$^1$- P.N. Lebedev Physical Institute of the Russian Academy of Sciences. Leninsky Prospekt 53, 119991, Moscow, Russia\\
$^2$- National Research University "MPEI", Krasnokazarmennaya, 14, Moscow 111250, Russia\\
$^3$- Moscow Institute of Physics and Technology, 141700 Dolgoprudny, Russia
}%
\email{alexkun@lebedev.ru}

\date{\today}

\begin{abstract}
While much effort are now concentrated on exploration of crystalline form of MoS$_2$ as a two-dimensional material,  we show that room-temperature DC magnetron sputtering from pressurized MoS$_2$ powder target produces amorphous MoS$_x$ films with optical properties fundamentally different from those of crystalline MoS$_2$. The films were found to have a sulfur content of approximately $x=1.8$, indicating sulfur deficiency. They are optically isotropic and exhibit rather low optical losses, with a refractive index of about 2 over the visible and near-infrared spectral range. Their compatibility with lift-off lithography and dry etching makes amorphous MoS$_x$ a promising material for integrated photonic applications.
\end{abstract}

\maketitle

\section{Introduction}
Transition metal dichalcogenides (TMDCs) are remarkable for their outstanding optical properties\cite{manzeli20172d} and stable under ambient conditions\cite{longo2017intrinsic}. MoS$_2$ is one of the most widely used and cheapest TMDCs\cite{li2015two}. High-quality bulk crystals occur naturally as the mineral molybdenite. Powdered MoS$_2$ is widely used as a lubricant and catalyst component\cite{merki2011amorphous}. Crystalline monolayers and poly-crystalline films sparkle nanoelectronic developments: flexible electronics\cite{singh2019flexible,hoang2023low}, valleytronics\cite{mak2014valley, schaibley2016valleytronics} and ultra-scaled transistors\cite{wu2022vertical}. Thick TMDC films have also attracted attention for memristive devices\cite{li2018mos2}, optoelectronics and nanophotonics because of their high refractive index\cite{ling2021all, muhammad2021optical, zotev2025nanophotonics}.

Optical applications of MoS$_2$ require reproducibility, scalability and therefore cannot rely on individually exfoliated flakes. There are many methods for the growth of MoS$_2$ films: liquid exfoliation, molecular beam epitaxy, oxide sulfurization, pulsed laser deposition\cite{wang2017pulsed} ,metal-organic chemical vapor deposition, atomic layer deposition\cite{hamtaei2025direct}, physical vapour transport, spray pyrolysis, sol-gel techniques, see Ref. \cite{sun2017synthesis} for review. Magnetron sputtering is among the most cost-effective deposition techniques. It can produce uniform films without the use of toxic components and ultra-high vacuum. Previous studies of magnetron-sputtered MoS$_2$ films have mainly focused either on tribological properties\cite{qin2013microstructure, gu2015amorphous, ren2019structural, lu2021novel, lu2021exploring} or on 2D limit\cite{ling2015large, samassekou2017viable}.

The potential of TMDC films with thicknesses of several hundred nanometers for optical and photonic applications remains largely unexplored. In this paper we demonstrate the fabrication of amorphous magnetron-sputtered MoS$_x$ films and investigate their structural, chemical, and optical properties. The films exhibit a sulfur-deficient composition close to MoS$_{1.8}$, good uniformity, and long-term stability under ambient conditions. Surprisingly, the films are almost electrically insulating and show much smaller optical extinction than expected for a doped semiconductor. Their optical properties are isotropic and also differ from the bulk MoS$_2$ crystal. The combination of the obtained optical properties and compatibility with standard microfabrication technologies make amorphous MoS$_x$ a promising material for future photonic applications.

\section{Methods}
\subsection{Sample preparation}

\begin{figure}[htbp]
    \centering
    \includegraphics[width=\linewidth]{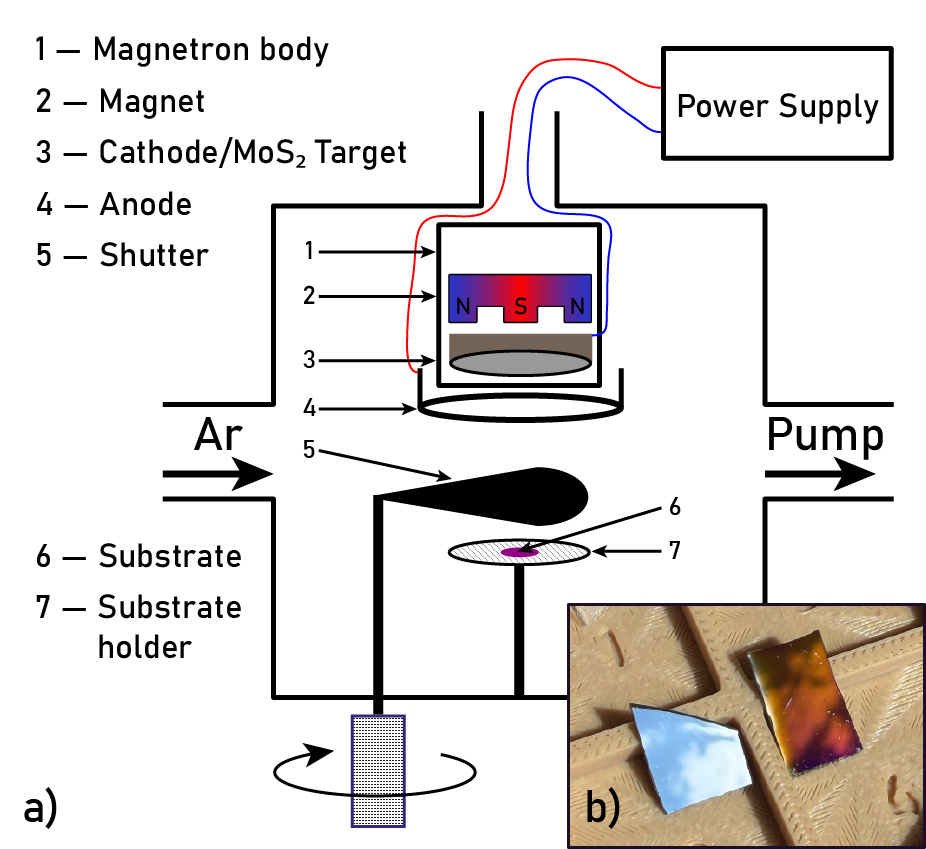}
    \caption{a) Home-built DC magnetron sputtering system for MoS$_x$ film deposition. b) Photograph of a Si substrate (on the left) and a Si substrate coated with a 25 nm
MoS$_x$ film (on the right).}
    \label{fig:Mag}
\end{figure}

Commercial 98.5\% purity MoS$_2$ Molykote Z\texttrademark{} Powder was pressed in an aluminum mold to form a target 38 mm in diameter and approximately 1 mm thick at a pressure of about 30 MPa. The pressed target was annealed in vacuum. The annealing procedure consisted of heating to 1000 \celsius~over 30 min, followed by a 3 hour dwell at the maximum temperature and subsequent slow cooling to room temperature. During annealing, the quartz tube was continuously evacuated to a pressure below $10^{-3}$ mbar.

Film deposition was carried out using a DC magnetron installed in a home-built vacuum chamber (Fig.~\ref{fig:Mag}) in an argon atmosphere at a pressure of $(6\text{--}7)\times10^{-2}$ mbar. The target-to-substrate distance was 10 cm, and the discharge power was $\approx$100 W. Since the magnetron was operated without water cooling, sputtering was performed in 1 min deposition cycles separated by 1 min cooling intervals to prevent overheating of the target and magnetron. The deposition rate of $\approx$0.3 nm/s was determined from ex-situ atomic force microscopy (AFM) thickness measurements.

The films were deposited on silicon, sapphire, and glass substrates depending on the particular experiment. Unless otherwise stated, the structural and chemical characterization presented in this work was performed on films deposited on Si substrates.

Bulk crystalline 2H-MoS$_2$ (natural molybdenite) was used as a reference material for structural, optical, and XPS measurements. The crystal structure of the fabricated MoS$_2$ target was verified by powder X-ray diffraction (XRD) before and after annealing. Figure~\ref{fig:XRD}(a) shows the diffraction patterns of the target before and after annealing together with that of the reference MoS$_2$ single crystal. The characteristic reflections of the 2H-MoS$_2$ phase, including (002), (103), (006), (105), and (110), are clearly observed. After annealing, the target exhibits a larger number of well-defined diffraction peaks, indicating improved structural ordering.


\subsection{Characterization techniques}

Film thickness was determined from the height of a step created by locally removing the film with a sharp needle. Surface topography was studied using an NT-MDT Solver atomic force microscope operating in semi-contact mode.

XRD measurements were performed using a Rigaku SmartLab diffractometer equipped with a Cu K$\alpha$ X-ray source ($\lambda = 1.5406$ Å, Siemens KFL CU 2k).

Chemical composition and bonding states were investigated by X-ray photoelectron spectroscopy (XPS) using a NanoFab 25 electron-ion spectroscopy platform equipped with a PHOIBOS HSA3500 CCD 225 R6-HiRes electron energy analyzer. An XR-50 X-ray source with an Al anode was used for excitation. Survey spectra were acquired with a pass energy of 80 eV, while high-resolution spectra were recorded with a pass energy of 20 eV.


Spectral reflectance and transmittance measurements were performed over the range from 400 to 1500 nm using a J.A. Woollam VASE ellipsometer. In order to find MoS$_x$ refractive ($n_f$) and absorption ($k_f$) indices from the experimentally measured ellipticity parameters, the samples were modeled as a substrate, coated with a uniform film, and the system of equations was solved numerically:
\begin{equation}
\Psi(n_f, k_f, d_f, n_s, k_s, \alpha) = \Psi_{\rm{exp}}
\end{equation}
\begin{equation}
\Delta(n_f, k_f, d_f, n_s, k_s, \alpha) = \Delta_{\rm{exp}}
\end{equation}
Here $\Psi$ and $\Delta$ are ellipsometric parameters, $\alpha$ is the incidence angle, $d$ is the film thickness, and $n_s$, $k_s$ are the indices of refraction and absorption of the substrate, respectively. We chose a physically plausible solution with a continuous smooth
spectrum which satisfies the requirements $n, k > 0$.

Electrical measurements were performed using a Keithley 6514 electrometer in a two-point probe configuration.

\section{Results}

\subsection{Structural properties}
\begin{figure}[htbp]
    \centering
    \includegraphics[width=1\linewidth]{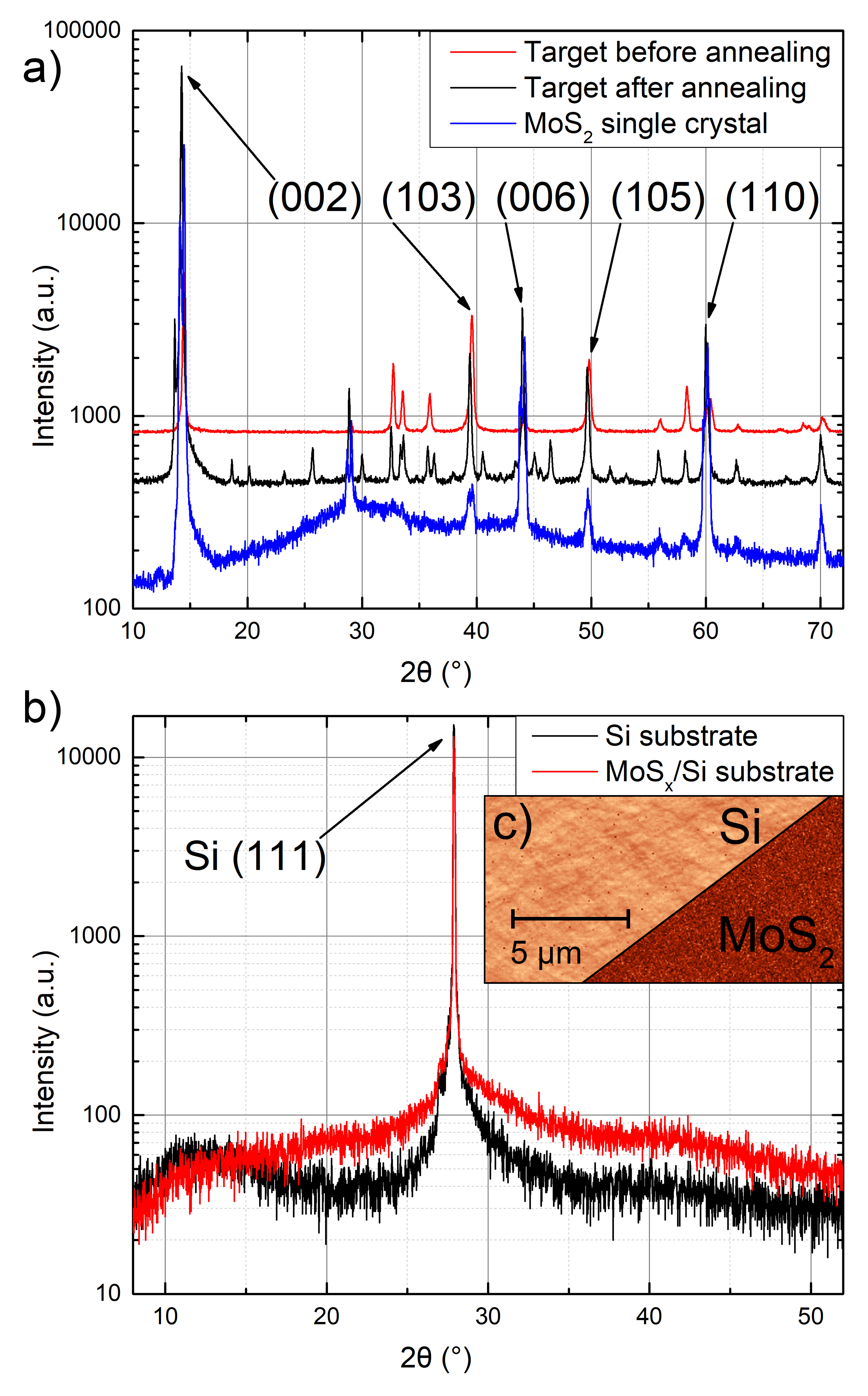}
    \caption{X-ray diffraction patterns of the investigated samples. (a) Diffraction patterns of the pressed MoS$_2$ target before annealing, the annealed target, and a bulk 2H-MoS$_2$ single crystal used as a reference. The main reflections of the 2H-MoS$_2$ phase are indexed. The patterns are vertically offset by 300 a.u. for clarity.(b) Comparison of the bare Si substrate and the sputtered MoS$_x$/Si sample. Only the Si(111) substrate reflection is observed for the coated sample. No diffraction peaks corresponding to crystalline MoS$_2$ were detected. The intensity axes in both panels are shown on a logarithmic scale. (c) AFM topography images of the bare Si substrate (upper left) and the sputtered MoS$_x$ film (lower right). Both images are displayed using the same height scale.}
    \label{fig:XRD}
\end{figure}

Figure \ref{fig:XRD}(b) compares the XRD patterns of the bare Si substrate and the sputtered MoS$_x$/Si sample. The coated sample exhibits only the Si(111) substrate reflection at $2\theta \approx 28^\circ$. No diffraction peaks corresponding to crystalline MoS$_2$ were detected, indicating that the deposited MoS$_x$ film is amorphous.


{Figure~\ref{fig:XRD}(c) shows the AFM topography of the bare Si substrate and the sputtered MoS$_x$ film.} Surface roughness was characterized by the arithmetic average roughness ($S_a$) and the root mean square roughness ($S_q$). The bare silicon substrate exhibited $S_a = 0.23$ nm and $S_q = 0.33$ nm. After deposition, the roughness increased to $S_a = 1.7$  nm and $S_q = 2.1$ nm. No grains or faceted crystallites were observed, consistent with the amorphous structure indicated by the XRD results.

\subsection{Electrical properties}
Electrical measurements revealed sheet resistances exceeding 200 M$\Omega/\square$ for 30 nm thick film at room temperature. This value was independent of ambient illumination presence in striking contrast with crystalline MoS$_2$, remarkable by photoconductivity effects\cite{mak2014valley}. These values indicate that the sputtered films show insulating behavior and weakly interact with visible light.

This result is particularly noteworthy because sulfur-deficient amorphous MoS$_x$ films prepared by RF magnetron sputtering are often reported to exhibit measurable electrical conductivity \cite{krbal2023anomalous,huang2019amorphous,alev2025effects}. Despite the sulfur deficiency revealed by XPS (approximately MoS$_{1.8}$, see below), our films remain electrically insulating.


\begin{figure}[ht]
    \centering
    \includegraphics[width=\linewidth]{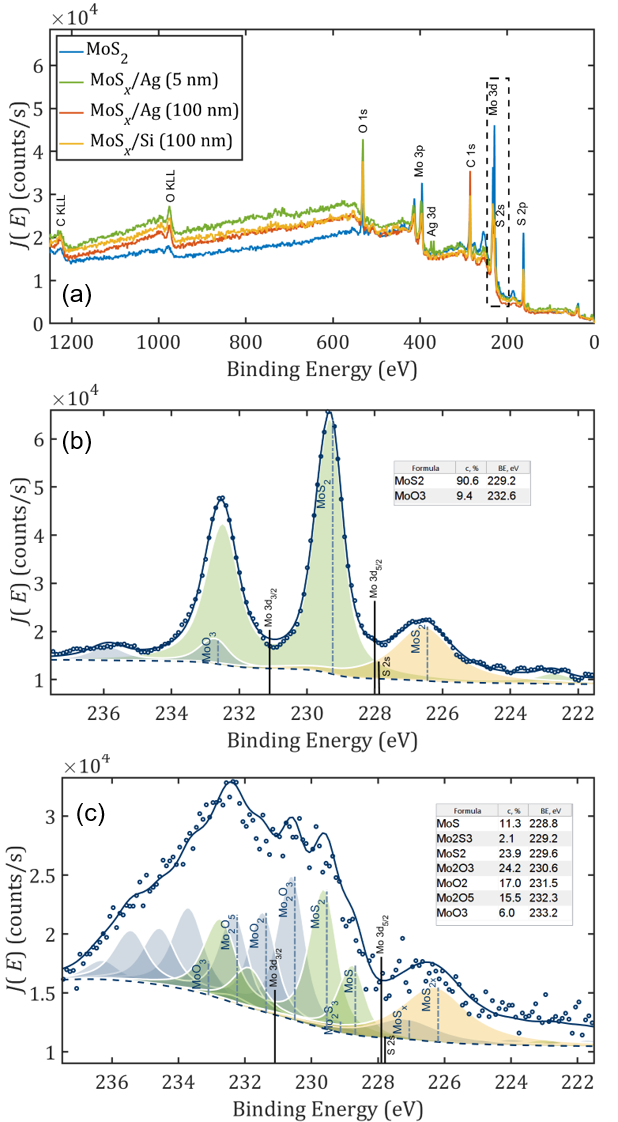}
    \caption{XPS spectra of the studied films compared to bulk MoS$_2$. (a) Review spectra; (b) molybdenite crystal (Mo peak) and its decomposition onto MoS$_2$ and MoO$_3$; (b) example of the amorphous film spectrum and its decomposition into series of oxides and sulfides.}
    \label{fig:XPS1}
\end{figure}

\subsection{Chemical composition (XPS)}
Figure~\ref{fig:XPS1} presents the XPS results obtained for the sputtered MoS$_x$ films and the reference bulk crystalline 2H-MoS$_2$. The survey spectra are shown in Fig.~\ref{fig:XPS1}a, while the Mo 3$d$ core-level spectra together with the corresponding peak fitting are presented in Fig.~\ref{fig:XPS1}b,c for the crystalline reference and a representative sputtered film, respectively. Comparison of the spectra reveals that, unlike the nearly stoichiometric bulk crystal, which contains only a minor contribution from molybdenum oxide, the sputtered films exhibit multiple molybdenum oxidation states together with sulfur in different chemical environments.

Depth profiling was performed using a method described in \cite{lub2018}. The quantitative layer model obtained from depth profiling is summarized in Table~\ref{tab:XPS}. For all sputtered films, the analysis reveals a thin surface oxide layer approximately 2--3 nm thick covering a sulfur-deficient MoS$_x$ layer. Such surface oxidation is consistent with the well-known oxygen-for-sulfur substitution in molybdenum sulfides \newt{\cite{1987_Buck,krbal2021}}. Despite different substrates and film thicknesses, the chemical composition of the sulfide layer remains remarkably similar, corresponding approximately to MoS$_{1.8}$. 

\begin{table}[ht]
\caption{Composition of the surface oxide layer determined from XPS depth profiling before and after two weeks of exposure to ambient air.}
\label{tab:oxidation}
\centering
\begin{tabular}{lcc}
\toprule
Component & Fresh film (\%) & After 2 weeks in air(\%) \\
\midrule
MoO$_3$ & 34 & 55 \\
MoO$_2$ & 36 & 28 \\
Mo$_2$O$_5$ & 30 & 17 \\
Oxide thickness (nm) & $2.0 \pm 0.3$ & $2.5 \pm 0.3$ \\
\bottomrule
\end{tabular}
\end{table}

\begin{figure}[ht]
    \centering
    \includegraphics[width=\linewidth]{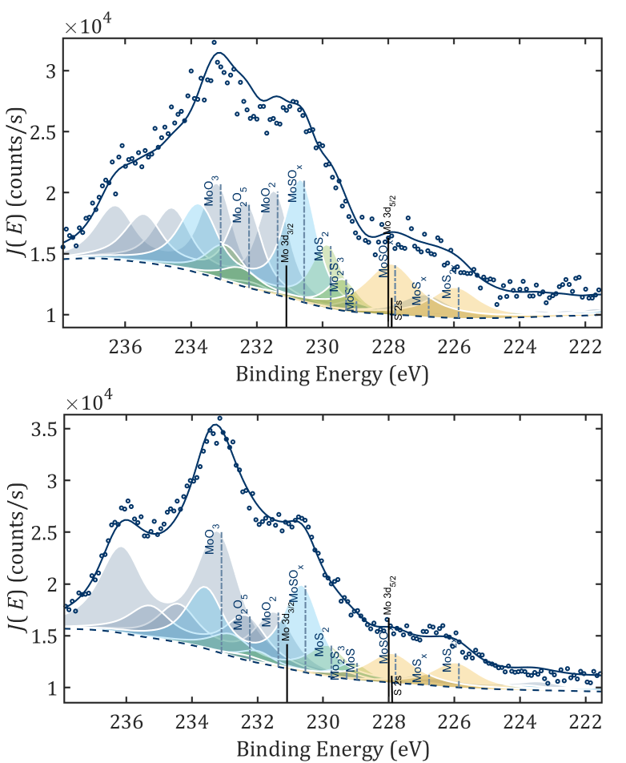}
    \caption{Comparison of the Mo 3$d$ XPS spectra of the same film measured immediately after deposition (top) and after two weeks of exposure to ambient air (bottom).}
    \label{fig:XPS2}
\end{figure}

To investigate the oxidation dynamics, the same film was analyzed by XPS immediately after deposition and after two weeks of exposure to ambient air. The corresponding spectra are shown in Fig.~\ref{fig:XPS2}, while the results of the quantitative compositional analysis are summarized in Table~\ref{tab:oxidation}. The analysis indicates that the oxide layer increased from $2.0 \pm 0.3$ nm to $2.5 \pm 0.3$ nm, accompanied by an increase in the relative fraction of MoO$_3$ and a corresponding decrease in the fractions of MoO$_2$ and Mo$_2$O$_5$. These results suggest an oxidation rate of approximately 0.3 nm per week under ambient conditions. The observed oxidation rate indicates that the films remain sufficiently stable for ex-situ handling and optical characterization.

\begin{table*}[ht]
\caption{Layer model derived from XPS depth profiling}
\label{tab:XPS}
\centering
\begin{tabular}{llll}
\toprule
Sample & Layer & Thickness (nm) & Composition \\
\midrule

\multirow{2}{*}{Bulk 2H-MoS$_2$}
& Surface & $0.23\pm0.05$ & MoO$_3$ \\
& Bulk & $\infty$ & MoS$_2$ \\
\midrule

\multirow{3}{*}{MoS$_x$/Ag (5 nm)}
& Surface oxide & $3.0\pm0.4$
& Mo$_2$O$_5$ (25\%), MoO$_2$ (27\%), Mo$_2$O$_3$ (39\%), MoO$_3$ (9\%) \\
& Film & $5.4\pm0.4$
& MoS (30\%), Mo$_2$S$_3$ (6\%), MoS$_2$ (64\%) \\
& Substrate & $\infty$ & Ag \\
\midrule

\multirow{2}{*}{MoS$_x$/Ag (100 nm)}
& Surface oxide & $2.2\pm0.3$
& Mo$_2$O$_5$ (27\%), MoO$_2$ (21\%), Mo$_2$O$_3$ (32\%), MoO$_3$ (20\%) \\
& Film & $\infty$
& MoS (33\%), Mo$_2$S$_3$ (11\%), MoS$_2$ (56\%) \\
\midrule

\multirow{2}{*}{MoS$_x$/Si (100 nm)}
& Surface oxide & $2.3\pm0.3$
& Mo$_2$O$_5$ (33\%), MoO$_2$ (19\%), Mo$_2$O$_3$ (35\%), MoO$_3$ (13\%) \\
& Film & $\infty$
& MoS (38\%), Mo$_2$S$_3$ (7\%), MoS$_2$ (55\%) \\
\bottomrule
\end{tabular}
\end{table*}

\subsection{Optical constants (ellipsometry)}

\begin{figure}[ht]
    \centering
    \includegraphics[width=0.8\linewidth]{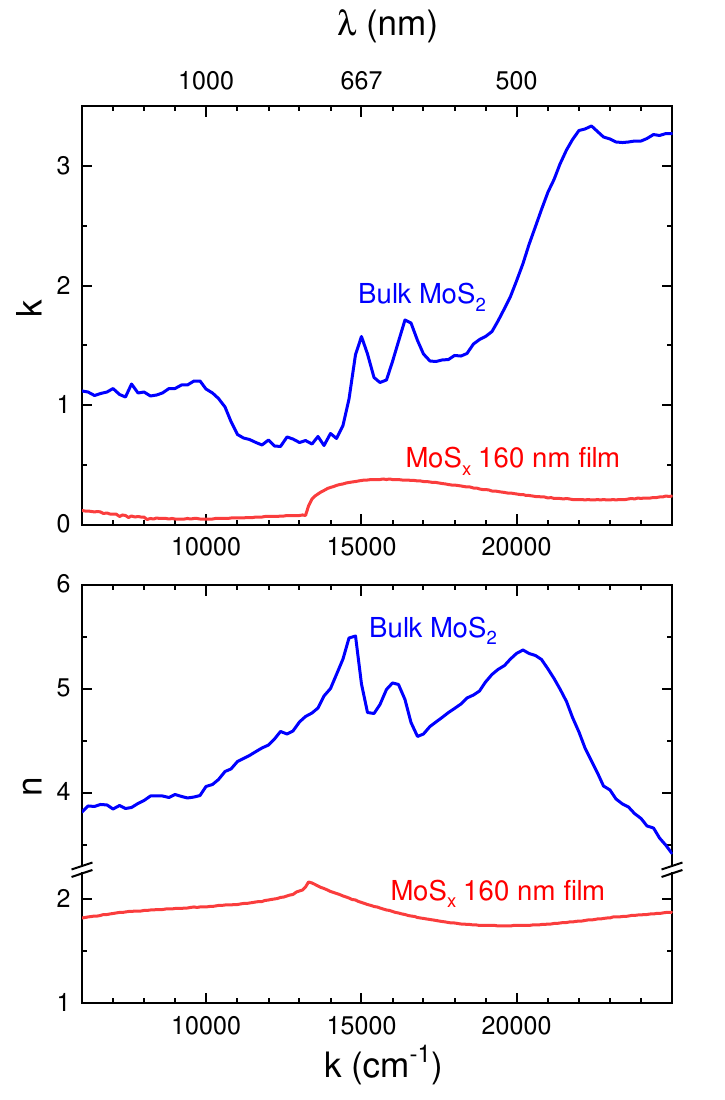}
    \caption{Optical constants of the sputtered 160 nm MoS$_x$ film and bulk 2H-MoS$_2$: (a) refractive index $n$, (b) extinction coefficient $k$ as a function of wavelength.}
    \label{fig:Ellipsometry}
\end{figure}

Figure~\ref{fig:Ellipsometry} compares the optical constants of bulk 2H-MoS$_2$ and the sputtered MoS$_x$ film. Two pronounced absorption resonances are observed near 15000 and 16500 cm$^{-1}$, corresponding to the characteristic excitonic transitions of crystalline MoS$_2$.


The optical constants of the sputtered films differ markedly from those of bulk crystalline MoS$_2$. The refractive index remains close to 1.6–2.1 over most of the investigated spectral range, whereas crystalline MoS$_2$ exhibits values exceeding 4 near the excitonic resonances. Likewise, the characteristic absorption peaks observed in the single crystal are absent in the sputtered films.

These results indicate that optical properties of the sputtered films differ fundamentally from crystalline 2H-MoS$_2$.

\section{Discussion}
In order to obtain an amorphous film, it is essential to maintain the substrate at a sufficiently low temperature so that crystalline or polycrystalline phases do not form. Substrate heating during sputtering is known to promote the formation of more crystalline MoS$_2$, whereas the main focus of the present study is the achievement of an electrically insulating amorphous state.

There are several reports of amorphous MoS$_2$ films with rather different properties\cite{1987_Buck, merki2011amorphous, wang2017pulsed, krbal2023anomalous} obtained by different methods: some of them are conductive\cite{krbal2023anomalous}, some of them are poorly conductive \cite{shin1988}; some of them contain the presence of MoS$_2$ crystalline phase MoS$_2$\cite{1998_McDevitt}, some are not\cite{gellerup2023room}. These parameters depend on the stoichiometry and the structure of the films.
 
Our work demonstrates reproducible and insulating sulfur-deficient films. This deficiency comes from the sulfur loss during annealing of the pressed MoS$_2$ target and sputtering itself. Interestingly, several reports using alternative deposition routes have obtained nearly stoichiometric or even sulfur-rich amorphous MoS$_x$ films\cite{Sulfur_deficient_films}.

We believe the physical reason of the insulating character is that crystalline molybdenium sulfides MoS$_x$ with $x<2$ (MoS\cite{WEI2017114}, Mo$_2$S$_3$\cite{panchu2020neodymium}, Mo$_6$S$_8$\cite{popov2007structural}, etc ) are semiconducting. Band gap further increases in the nano-structured grains within the film and therefore resonant optical features are smeared. High conductivity of the amorphous MoS$_x$ films in Ref.~\cite{krbal2023anomalous} was attributed to Mo-Mo homopolar bonds that exceed the percolation threshold. In our films the concentration of such bonds is apparently below the threshold. 

Low dc magnetron power allowed us to avoid its water cooling. We especially highlight minimal requirements needed to fabricate the films. The target is produced from the extremely cheap and mass-producted MoS$_2$ powder, without high hydrostatic pressure. Hardware simplicity makes the fabrication of MoS$_x$ films accessible even in laboratories lacking dedicated thin-film deposition infrastructure.

The deposition was carried out without intentional substrate heating, and we checked the compatibility of MoS$_x$ film technology with standard optical and electron photoresist lift-off, that also evidences for absence of strong heating. This compatibility opens a pathway to photonic devices: waveguides, resonators, beam splitters. 
 

Amorphous MoS$_x$ and lithographically pre-patterned structures made out of them also could serve as a starting point for further functionalization, e.g. by chemical, thermal or laser/electron beam treatment. For example Refs.\cite{krbal2021,ocana2024measuring} demonstrate recrystallization of amorphous MoS$_2$ films under heating (see Ref. \cite{krbal2026tunability}). It demonstrates laser-induced change of MoS$_2$ properties, Ref. \cite{2017_Kim} shows electron-beam induced recrystallization.
Our work thus opens a pathway to some possible inexpensive MoS$_2$-based devices.

\section{Conclusions}

We demonstrated a simple, low-cost, and reproducible approach for the deposition of amorphous MoS$_2$ thin films by magnetron sputtering. The deposited films remain electrically insulating, which makes them potentially useful in photonic and related devices and suggests a prominent example of disorder-induced insulating behavior in chalcogenides. At the same time, the amorphous state and compatibility with lift-off processes provides a convenient starting point for further controlled recrystallization, opening a route toward post-growth engineering of the structural and electronic properties of MoS$_2$ films.


\begin{acknowledgments}
Acknowledgments. 
 This work was supported by the Russian Science Foundation under Grant No. 25‑79‑20031 for magnetron sputtering of a protective MoS$_2$ film for an imprint stamp.
  Sample fabrication and studies was conducted at the Shared Facility Center of the P.N. Lebedev Physical Institute. 
 
\end{acknowledgments}

\nocite{*}
\bibliography{aipsamp}

@PREAMBLE{
 "\providecommand{\noopsort}[1]{}" 
 # "\providecommand{\singleletter}[1]{#1}%" 
}

@article{manzeli20172d,
  title={2D transition metal dichalcogenides},
  author={Manzeli, Sajedeh and Ovchinnikov, Dmitry and Pasquier, Diego and Yazyev, Oleg V and Kis, Andras},
  journal={Nature Reviews Materials},
  volume={2},
  number={8},
  pages={17033},
  year={2017},
  publisher={Nature Publishing Group}
}

@article{mak2014valley,
  title={The valley Hall effect in MoS2 transistors},
  author={Mak, Kin Fai and McGill, Kathryn L and Park, Jiwoong and McEuen, Paul L},
  journal={Science},
  volume={344},
  number={6191},
  pages={1489--1492},
  year={2014},
  publisher={American Association for the Advancement of Science}
}

@article{gellerup2023room,
  title={Room temperature magnetron sputtering and laser annealing of ultrathin amorphous sulfur-rich MoSx films},
  author={Gellerup, Spencer and Arnold, Corey L and Muratore, Christopher and Glavin, Nicholas R and Shepherd, Nigel D and Voevodin, Andrey A},
  journal={Journal of Vacuum Science \& Technology A},
  volume={41},
  number={5},
  year={2023},
  publisher={AIP Publishing}
}

@article{alev2025effects,
  title={Effects of target power and deposition pressure on magnetron-sputtered molybdenum disulfide thin films: Morphological, structural, optical, and electrical characteristics},
  author={Alev, Onur and {\"O}zdemir, Okan and K{\i}l{\i}{\c{c}}, Alp and B{\"u}y{\"u}kk{\"o}se, Serkan and Goldenberg, Eda},
  journal={Ceramics International},
  volume={51},
  number={7},
  pages={8607--8614},
  year={2025},
  publisher={Elsevier}
}

@article{schaibley2016valleytronics,
  title={Valleytronics in 2D materials},
  author={Schaibley, John R and Yu, Hongyi and Clark, Genevieve and Rivera, Pasqual and Ross, Jason S and Seyler, Kyle L and Yao, Wang and Xu, Xiaodong},
  journal={Nature Reviews Materials},
  volume={1},
  number={11},
  pages={16055},
  year={2016},
  publisher={Nature Publishing Group}
}

@article{merki2011amorphous,
  title={Amorphous molybdenum sulfide films as catalysts for electrochemical hydrogen production in water},
  author={Merki, Daniel and Fierro, St{\'e}phane and Vrubel, Heron and Hu, Xile},
  journal={Chemical Science},
  volume={2},
  number={7},
  pages={1262--1267},
  year={2011},
  publisher={The Royal Society of Chemistry}
}

@article{popov2007structural,
  title={Structural and electronic properties of Mo 6 S 8 clusters deposited on a Au (111) surface investigated with density functional theory},
  author={Popov, Igor and Gemming, Sibylle and Seifert, Gotthard},
  journal={Physical Review B—Condensed Matter and Materials Physics},
  volume={75},
  number={24},
  pages={245436},
  year={2007},
  publisher={APS}
}

@article{wang2017pulsed,
  title={Pulsed laser deposition of amorphous molybdenum disulfide films for efficient hydrogen evolution reaction},
  author={Wang, Ruijing and Sun, Peng and Wang, Huanwen and Wang, Xuefeng},
  journal={Electrochimica Acta},
  volume={258},
  pages={876--882},
  year={2017},
  publisher={Elsevier}
}

@inproceedings{hamtaei2025direct,
  title={Direct ALD of amorphous MoS2 thin films for extra-terrestrial photovoltaic applications},
  author={Hamtaei, Sarallah and Kim, Sungjoon and Nazif, Koosha Nassiri and Nattoo, Crystal and Carr, Joshua M and Romanetz, Leo and Nitta, Frederick U and Reid, Obadia G and Vermang, Bart and Elam, Jeffery and others},
  booktitle={2025 IEEE 53rd Photovoltaic Specialists Conference (PVSC)},
  pages={1472--1472},
  year={2025},
  organization={IEEE}
}

@article{panchu2020neodymium,
  title={Neodymium YAG laser chemical vapor deposition growth of luminescent Mo2S3 nanocrystals using bulk MoS2 and its structural, optical properties and caspase-mediated apoptosis in THP-1 monocytic cells},
  author={Panchu, SJ and Dhani, S and Chuturgoon, AA and Swart, HC and Moodley, MK},
  journal={Materials Today Chemistry},
  volume={17},
  pages={100315},
  year={2020},
  publisher={Elsevier}
}

@article{WEI2017114,
title = {Prediction of stable ground-state and pressure-induced phase transition of molybdenum monosulfide},
journal = {Materials Science and Engineering: B},
volume = {226},
pages = {114-119},
year = {2017},
issn = {0921-5107},
doi = {https://doi.org/10.1016/j.mseb.2017.09.013},
url = {https://www.sciencedirect.com/science/article/pii/S0921510717302301},
author = {Qun Wei and Quan Zhang and Haiyan Yan and Meiguang Zhang and Xiaofeng Shi and Xuanmin Zhu},
}

@article{li2018mos2,
  title={MoS2 memristors exhibiting variable switching characteristics toward biorealistic synaptic emulation},
  author={Li, Da and Wu, Bin and Zhu, Xiaojian and Wang, Juntong and Ryu, Byunghoon and Lu, Wei D and Lu, Wei and Liang, Xiaogan},
  journal={ACS nano},
  volume={12},
  number={9},
  pages={9240--9252},
  year={2018},
  publisher={ACS Publications}
}

@article{DefectsHu2018two,
  title={Two-dimensional transition metal dichalcogenides: interface and defect engineering},
  author={Hu, Zehua and Wu, Zhangting and Han, Cheng and He, Jun and Ni, Zhenhua and Chen, Wei},
  journal={Chemical Society Reviews},
  volume={47},
  number={9},
  pages={3100--3128},
  year={2018},
  publisher={Royal Society of Chemistry}
}

@article{li2015two,
  title={Two-dimensional MoS2: Properties, preparation, and applications},
  author={Li, Xiao and Zhu, Hongwei},
  journal={Journal of Materiomics},
  volume={1},
  number={1},
  pages={33--44},
  year={2015},
  publisher={Elsevier}
}

@article{sun2017synthesis,
  title={Synthesis methods of two-dimensional MoS2: A brief review},
  author={Sun, Jie and Li, Xuejian and Guo, Weiling and Zhao, Miao and Fan, Xing and Dong, Yibo and Xu, Chen and Deng, Jun and Fu, Yifeng},
  journal={Crystals},
  volume={7},
  number={7},
  pages={198},
  year={2017},
  publisher={MDPI}
}

@article{qin2013microstructure,
  title={Microstructure, mechanical and tribological behaviors of MoS2-Ti composite coatings deposited by a hybrid HIPIMS method},
  author={Qin, Xiaopeng and Ke, Peiling and Wang, Aiying and Kim, Kwang Ho},
  journal={Surface and Coatings Technology},
  volume={228},
  pages={275--281},
  year={2013},
  publisher={Elsevier}
}

@article{lu2021novel,
  title={A novel design by constructing MoS2/WS2 multilayer film doped with tantalum toward superior friction performance in multiple environment},
  author={Lu, Zhaoxia and Zhang, Chaozhi and Zeng, Chun and Ren, Siming and Pu, Jibin},
  journal={Journal of Materials Science},
  volume={56},
  number={31},
  pages={17615--17631},
  year={2021},
  publisher={Springer}
}

@article{singh2019flexible,
  title={Flexible molybdenum disulfide (MoS2) atomic layers for wearable electronics and optoelectronics},
  author={Singh, Eric and Singh, Pragya and Kim, Ki Seok and Yeom, Geun Young and Nalwa, Hari Singh},
  journal={ACS applied materials \& interfaces},
  volume={11},
  number={12},
  pages={11061--11105},
  year={2019},
  publisher={ACS Publications}
}

@article{mooshammer2024enabling,
  title={Enabling waveguide optics in rhombohedral-stacked transition metal dichalcogenides with laser-patterned grating couplers},
  author={Mooshammer, Fabian and Xu, Xinyi and Trovatello, Chiara and Peng, Zhi Hao and Yang, Birui and Amontree, Jacob and Zhang, Shuai and Hone, James and Dean, Cory R and Schuck, P James and others},
  journal={ACS nano},
  volume={18},
  number={5},
  pages={4118--4130},
  year={2024},
  publisher={ACS Publications}
}

@article{zotev2025nanophotonics,
  title={Nanophotonics with multilayer van der Waals materials},
  author={Zotev, Panaiot G and Bouteyre, Paul and Wang, Yadong and Randerson, Sam A and Hu, Xuerong and Sortino, Luca and Wang, Yue and Shegai, Timur and Gong, Su-Hyun and Tittl, Andreas and others},
  journal={Nature Photonics},
  volume={19},
  number={8},
  pages={788--802},
  year={2025},
  publisher={Nature Publishing Group UK London}
}

@article{muhammad2021optical,
  title={Optical bound states in continuum in MoS2-based metasurface for directional light emission},
  author={Muhammad, Naseer and Chen, Yang and Qiu, Cheng-Wei and Wang, Guo Ping},
  journal={Nano Letters},
  volume={21},
  number={2},
  pages={967--972},
  year={2021},
  publisher={ACS Publications}
}

@article{ling2021all,
  title={All van der Waals integrated nanophotonics with bulk transition metal dichalcogenides},
  author={Ling, Haonan and Li, Renjie and Davoyan, Artur R},
  journal={Acs Photonics},
  volume={8},
  number={3},
  pages={721--730},
  year={2021},
  publisher={ACS Publications}
}

@article{longo2017intrinsic,
  title={Intrinsic air stability mechanisms of two-dimensional transition metal dichalcogenide surfaces: basal versus edge oxidation},
  author={Longo, Roberto C and Addou, Rafik and KC, Santosh and Noh, Ji-Young and Smyth, Christopher M and Barrera, Diego and Zhang, Chenxi and Hsu, Julia WP and Wallace, Robert M and Cho, Kyeongjae},
  journal={2D Materials},
  volume={4},
  number={2},
  pages={025050},
  year={2017},
  publisher={IOP Publishing}
}

@article{1998_McDevitt,
author = {N.T McDevitt and J. E. Bultman and J.S. Zabinski},
title = {Study of Amorphous MoS2 Films Grown by Pulsed Laser Deposition},
journal = {Applied Spectroscopy},
year = {1998},
volume = {52},
publisher = {SAGE},
month = {sep},
url = {https://doi.org/10.1366/0003702981945165},
number = {9},
pages = {1160--1164},
doi = {10.1366/0003702981945165}
}

@article{ocana2024measuring,
  title={Measuring the multifunctional properties of MoS 2 across the amorphous-crystalline transition using colorimetric sensing},
  author={Ocana-Pujol, Jose L and Gallivan, Rebecca A and Ord{\'o}{\~n}ez, Ram{\'o}n Camilo Dom{\'\i}nguez and Porenta, Nikolaus and M{\"u}ller, Arnold and Vockenhuber, Christof and Spolenak, Ralph and Galinski, Henning},
  journal={Physical Review B},
  volume={110},
  number={11},
  pages={115303},
  year={2024},
  publisher={APS}
}

@article{krbal2026tunability,
  title={Tunability of Amorphous MoS2 Thin Film Properties Through Pulsed KrF Laser Deposition Rate},
  author={Krbal, Milos and Prikryl, Jan and Prokop, Vit and Pereira, Jhonatan Rodriguez and Slang, Stanislav and Mistrik, Jan and Pis, Igor},
  journal={ACS Applied Materials \& Interfaces},
  year={2026},
 volume={18},
  number={12},
 pages = {18111--18119},
  publisher={ACS Publications}
}

@article{huang2019amorphous,
  title={Amorphous MoS2 photodetector with ultra-broadband response},
  author={Huang, Zhongzheng and Zhang, Tianfu and Liu, Junku and Zhang, Lihui and Jin, Yuanhao and Wang, Jiaping and Jiang, Kaili and Fan, Shoushan and Li, Qunqing},
  journal={ACS Applied Electronic Materials},
  volume={1},
  number={7},
  pages={1314--1321},
  year={2019},
  publisher={ACS Publications}
}

@article{jagosz2024wafer,
  title={Wafer-Scale Demonstration of Polycrystalline MoS2 Growth on 200 mm Glass and SiO2/Si Substrates by Plasma-Enhanced Atomic Layer Deposition},
  author={Jagosz, Julia and Willeke, Leander and Gerke, Nils and Becher, Malte JMJ and Plate, Paul and Kostka, Aleksander and Rogalla, Detlef and Ostendorf, Andreas and Bock, Claudia},
  journal={Advanced Materials Technologies},
  volume={9},
  number={22},
  pages={2400492},
  year={2024},
  publisher={Wiley Online Library}
}

@article{2017_Kim,
author = {Bong Ho Kim and Hyun Ho Gu and Young Joon Yoon},
title = {Atomic rearrangement of a sputtered MoS2 film from amorphous to a 2D layered structure by electron beam irradiation},
journal = {Scientific Reports},
year = {2017},
volume = {7},
publisher = {Springer Nature},
month = {jun},
url = {https://doi.org/10.1038/s41598-017-04222-6},
number = {1},
pages = {3874},
doi = {10.1038/s41598-017-04222-6}
}

@article{1987_Buck,
author = {Volker Buck},
title = {Preparation and properties of different types of sputtered MoS2 films},
journal = {Wear},
year = {1987},
volume = {114},
publisher = {Elsevier},
month = {feb},
url = {https://doi.org/10.1016/0043-1648(87)90116-5},
number = {3},
pages = {263--274},
doi = {10.1016/0043-1648(87)90116-5}
}

@article{krbal2021,
author = {Krbal, Milos and Prokop, Vit and Kononov, Alexey A. and Pereira, Jhonatan Rodriguez and Mistrik, Jan and Kolobov, Alexander V. and Fons, Paul J. and Saito, Yuta and Hatayama, Shogo and Shuang, Yi and Sutou, Yuji and Rozhkov, Stepan A. and Stellhorn, Jens R. and Hayakawa, Shinjiro and Pis, Igor and Bondino, Federica},
title = {Amorphous-to-Crystal Transition in Quasi-Two-Dimensional MoS2: Implications for 2D Electronic Devices},
journal = {ACS Applied Nano Materials},
volume = {4},
number = {9},
pages = {8834-8844},
year = {2021},
doi = {10.1021/acsanm.1c01504},

URL = {     
        https://doi.org/10.1021/acsanm.1c01504 
},
eprint = { 
        https://doi.org/10.1021/acsanm.1c01504
    }
}

@article{krbal2023anomalous,
  title={Anomalous electrical conductivity change in MoS2 during the transition from the amorphous to crystalline phase},
  author={Krbal, Milos and Prikryl, Jan and Pis, Igor and Prokop, Vit and Pereira, Jhonatan Rodriguez and Kolobov, Alexander V},
  journal={Ceramics International},
  volume={49},
  number={2},
  pages={2619--2625},
  year={2023},
  publisher={Elsevier}
}

@article{wu2022vertical,
  title={Vertical MoS$_2$ transistors with sub-1-nm gate lengths},
  author={Wu, Fan and Tian, He and Shen, Yang and Hou, Zhan and Ren, Jie and Gou, Guangyang and Sun, Yabin and Yang, Yi and Ren, Tian-Ling},
  journal={Nature},
  volume={603},
  number={7900},
  pages={259--264},
  year={2022},
  publisher={Nature Publishing Group UK London}
}

@article{hoang2023low,
  title={Low-temperature growth of MoS2 on polymer and thin glass substrates for flexible electronics},
  author={Hoang, Anh Tuan and Hu, Luhing and Kim, Beom Jin and Van, Tran Thi Ngoc and Park, Kyeong Dae and Jeong, Yeonsu and Lee, Kihyun and Ji, Seunghyeon and Hong, Juyeong and Katiyar, Ajit Kumar and others},
  journal={Nature nanotechnology},
  volume={18},
  number={12},
  pages={1439--1447},
  year={2023},
  publisher={Nature Publishing Group UK London}
}

@article{gu2015amorphous,
  title={Amorphous self-lubricant MoS2-C sputtered coating with high hardness},
  author={Gu, Lei and Ke, Peiling and Zou, Yousheng and Li, Xiaowei and Wang, Aiying},
  journal={Applied surface science},
  volume={331},
  pages={66--71},
  year={2015},
  publisher={Elsevier}
}

@article{lu2021exploring,
  title={Exploring the atmospheric tribological properties of MoS2-(Cr, Nb, Ti, Al, V) composite coatings by high throughput preparation method},
  author={Lu, Xiaolong and Yan, Mingming and Yan, Zhen and Chen, Wenyuan and Sui, Xudong and Hao, Junying and Liu, Weimin},
  journal={Tribology International},
  volume={156},
  pages={106844},
  year={2021},
  publisher={Elsevier}
}

@article{ren2019structural,
  title={Structural design of MoS2-based coatings toward high humidity and wide temperature},
  author={Ren, Siming and Shang, Kedong and Cui, Mingjun and Wang, Liping and Pu, Jibin and Yi, Peiyun},
  journal={Journal of Materials Science},
  volume={54},
  number={18},
  pages={11889--11902},
  year={2019},
  publisher={Springer}
}

@article{ling2015large,
  title={Large-scale two-dimensional MoS2 photodetectors by magnetron sputtering},
  author={Ling, ZP and Yang, R and Chai, JW and Wang, SJ and Leong, WS and Tong, Y and Lei, D and Zhou, Q and Gong, X and Chi, DZ and others},
  journal={Optics express},
  volume={23},
  number={10},
  pages={13580--13586},
  year={2015},
  publisher={Optical Society of America}
}

@article{samassekou2017viable,
  title={Viable route towards large-area 2D MoS2 using magnetron sputtering},
  author={Samassekou, Hassana and Alkabsh, Asma and Wasala, Milinda and Eaton, Miller and Walber, Aaron and Walker, Andrew and Pitk{\"a}nen, Olli and Kordas, Krisztian and Talapatra, Saikat and Jayasekera, Thushari and others},
  journal={2D Materials},
  volume={4},
  number={2},
  pages={021002},
  year={2017},
  publisher={IOP Publishing}
}

@article{lub2018,
	title = {XPS study of multilayer multicomponent films},
    author = {Lubenchenko, Alexander V. and Batrakov, Alexander A. and Pavolotsky, Alexey B. and Lubenchenko, Olga I. and Ivanov, Dmitriy A.},
    journal = {Applied Surface Science},
	volume = {427},
	pages = {711 -- 721},
    doi = {https://doi.org/10.1016/j.apsusc.2017.07.256},
	year={2018},
    publisher={Elsevier}
}

@article{shin1988,
  title={Effect of niobium doping on the properties of molybdenum sulfides as cathode materials},
  author={Shin, H and Doerr, HJ and Deshpandey, C and Fuqua, P and Dunn, B and Bunshah, RF},
  journal={Surface and Coatings Technology},
  volume={36},
  pages={859--865},
  year={1988},
  publisher={Elsevier}
}

@article{Sulfur_deficient_films,
author = {Matsuzaki, Kazunari and Sasaki, Iwao and Sunahara, Kenji and Ikeda, Michiaki and Matsuda, Kenji},
year = {2017},
month = {09},
pages = {675-681},
title = {Effect of Sulfur to Molybdenum Composition Ratio on Friction and Wear Characteristics in Air of Sputtered Molybdenum Disulfide Films},
volume = {66},
journal = {Journal of the Society of Materials Science, Japan},
doi = {10.2472/jsms.66.675}
}

\end{document}